# SEGMENTED FOIL SEM GRIDS AT FERMILAB*

S. Kopp,[#] D. Indurthy, Ž. Pavlović, M. Proga, R. Zwaska, Department of Physics,
University of Texas, Austin, TX 78712, U.S.A.
B. Baller, S. Childress, R. Ford, D. Harris, C. Kendziora, C. Moore, G. Tassotto,
Fermilab, Batavia, IL 60510, U.S.A.

*Abstract*

We present recent beam data from a new design of a profile monitor for proton beams at Fermilab. The monitors, consisting of grids of segmented Ti foils 5μm thick, are secondary-electron emission monitors (SEM's). We review data on the device's precision on beam centroid position, beam width, and on beam loss associated with the SEM material placed in the beam.

## INTRODUCTION

The Booster Neutrino Beam (BNB) [1] takes 8 GeV protons from the Booster accelerator and the "Neutrinos at the Main Injector (NuMI)" [2] beam takes 120 GeV protons from the Main Injector (MI) accelerator. Both beam lines require several times $10^{20}$ protons on target (POT) per year. Such fluences place stringent criteria on invasive instrumentation like SEM grids.

Based on a design from CERN [3] we have designed a SEM consisting of Ti foils segmented at either 1.0mm or 0.5mm pitch. The foils are 5μm thick Titanium, and two planes of the segmented foils per SEM chamber provides both horizontal and vertical beam profiles. The foil SEM's provide several features over the Au-plated 75μm Ø W-wire SEM's [4] in use at Fermilab: (1) a factor 50-60 lower fractional beam loss (see below), which is important for reduced component activation or groundwater contamination; (2) greater longevity of Ti signal yield [5], as compared with W or Au-W, which degraded by 20% over the course of running the KTeV fixed-target experiment [4]; (3) a 'bayonnette'-style frame permitting insertion/retraction from the beam without interruption of operations; and (4) reduced calculated beam-heating from the high-intensity proton-pulses, which results in less sag of the wires/foils [6].

Foil SEM's have been installed in the NuMI beam line in 2004. Additional detectors are envisioned for the 8 GeV line to the Booster Neutrino Beam and for the 8 GeV transfer line between the Booster and MI in 2005.

## SEM DESIGN

The design of the foil SEM chambers has been described in Ref. [7]. A few details are given here. The chambers have two planes of segmented signal foils for *X* and *Y* profiles of the beam. The foil strips have accordion "springs" pressed into their ends near the support ceramics to provide tensioning and compensation for beam heating. To achieve $5\times10^{-6}$ fraction beam loss the 1.0mm pitch chambers have strips which are 0.75mm wide near the edges, but are narrowed to 0.15mm width over the 75mm aperture of the beam. The paddle has an open area for the "beam out" position. Each segmented signal plane is interspersed with a 2.5μm thick "bias foil" to which 100 Volts is applied, drawing away the secondary electrons from the signal planes.

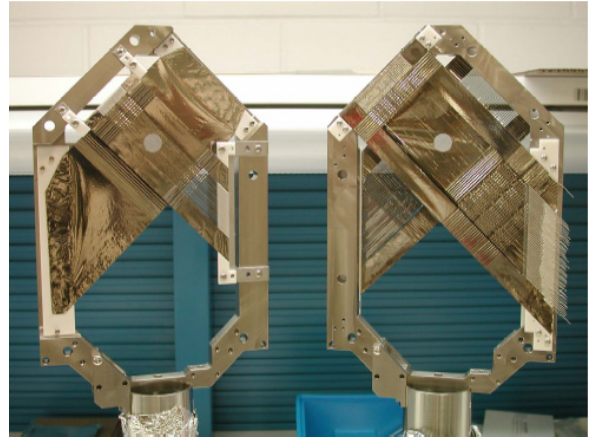

Figure 1: Shown at left (right) is a SEM paddle with strips at 0.5mm (1.0mm) pitch. Each wider exterior foils for measuring beam halo. The outer bias foils have been removed for clarity, but the middle bias foil is visible.

As described in Ref. [7], the actuation of the device is achieved by linear translation into the beam. The motion has ~4μm repeatability, with an additional 12μm long-term drift due to differential thermal expansion of the actuator. A linear variable differential transformer (LVDT) confirms the position.

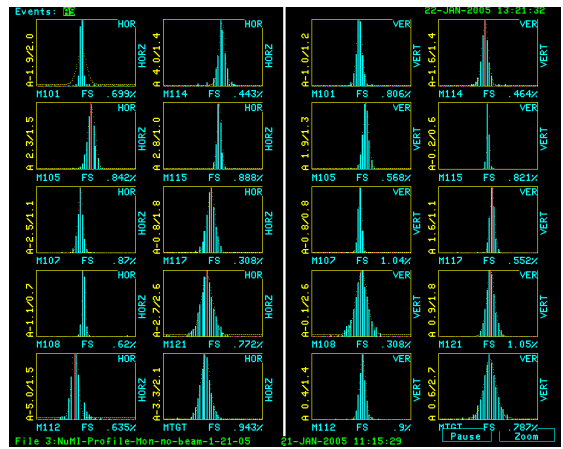

Figure 2: Beam profiles in the horizontal (left two columns) and vertical (right two columns) at 10 stations along the NuMI transport line, as measured by the foil SEM's. The last two stations are detectors with 0.5mm pitch for accurate extrapolation to the NuMI target.

___________________________________________
*Work supported by U.S. DoE, contracts DE-FG03-93ER40757 and DE-AC02-76CH3000
[#]kopp@hep.utexas.edu

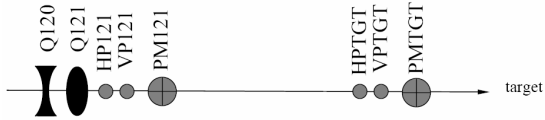

Figure 3: Schematic of the beam instrumentation upstream of the NuMI target, including two horizontal and two vertical BPM's, as well as two profile monitors.

Placement of the foil strips on the paddle was achieved with 10μm accuracy, as verified with optical measurements. The spring tensioning permits <40μm sag of the strips. Thermal simulations indicate that the linear expansion of the strips under beam heating of $4\times10^{13}$ ppp will be <10μm [6].

## NUMI BEAM EXPERIENCE

Figure 2 shows the beam profiles at 10 stations along the NuMI line. The SEM's have been used to commission the beam line, providing feedback and cross-checking of a new BPM system installed in the NuMI line, and confirming polarities of trims, dipoles, and quads.

In this section we present data on the SEM resolution of the beam centroid position and the beam width. Finally, we discuss losses seen as a result of the SEM foils placed in the beam. Studies of the resolutions are aided in two locations along the NuMI line where redundant instrumentation exists: in two locations are located regions of free drift in which there are two SEM's and two capacitive Beam Position Monitors (BPM's). The SEM's and BPM's are paired, with each pair separated by ~10m. In one location, the SEM's are 1.0mm pitch, while in the second region (shown in Figure 2) the SEM's are 0.5mm pitch.

Figure 3 shows the proton beam position as measured by the SEM labeled PMTGT and by the BPM labeled HPTGT during a scan across the NuMI target. The scatter about the diagonal is a consequence of the device resolutions. Inspection of Figure 4 shows horizontal bands as the beam position was incremented, indicating that the spread of measured positions at each increment is smaller along the SEM axis than it is along the BPM axis. At this beam intensity, the SEM resolution is apparently smaller than the BPM resolution.

Figure 5 shows the data of Figure 4 projected onto the BPM or the SEM axis. The number of spills within a particular bin of Figure 5 is indicative of the duration in time that the beam trajectory was set to a given location. The width of the peaks is indicative of the device resolutions. The fitted width of the peak at ~0.75mm, for example, is 18μm (54μm) for the SEM (BPM).

The observed SEM resolution of 18μm compares well to expectations from previous beam tests of the foil SEM's [8]. There, we measured the beam centroid resolution at three different beam widths. Extrapolating to the 1mm spot size in the NuMI beam, one expects 20-30μm resolution, consistent with the present observation.

Figure 6 shows the beam width as measured by PM121 and PMTGT over several hours. The beam converges as it heads to the NuMI target, but clearly the observed beam widths at the two stations correlate well. The beam width is typically ~0.9mm at PMTGT, but some variation exists spill-to-spill because of the variation in beam emittance from the Booster. The correlation between PM121 and PMTGT shows that the variation is not due to device resolution. Figure 7 shows a histogram of the beam widths measured by PMTGT over the same period (open histogram), as well as for a ~½ hour subset (shaded histogram). The shaded subset shows some of the same large-emittance spills, but also shows that a predominant number of spills at 0.93mm beam width. The fitted spread of the shaded peak is 23μm, indicating that the SEM's beam width resolution is smaller than this value. These observations are consistent with our beam test, which indicated a beam width resolution of 15-20μm for a beam width of twice the SEM grid pitch.

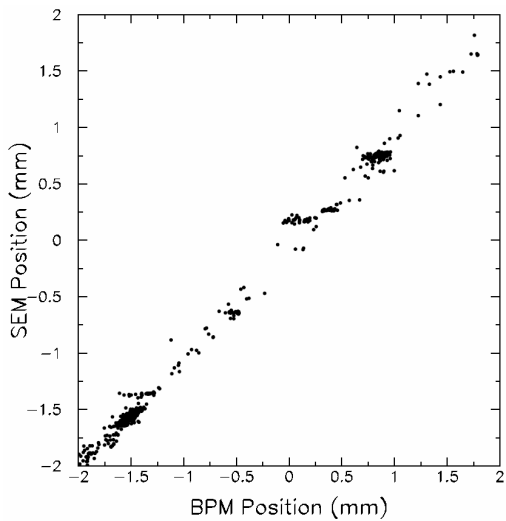

Figure 4: Horizontal beam centroid position as measured by SEM # PMTGT and the adjacent BPM # HPTGT during a portion of the NuMI commissioning run.

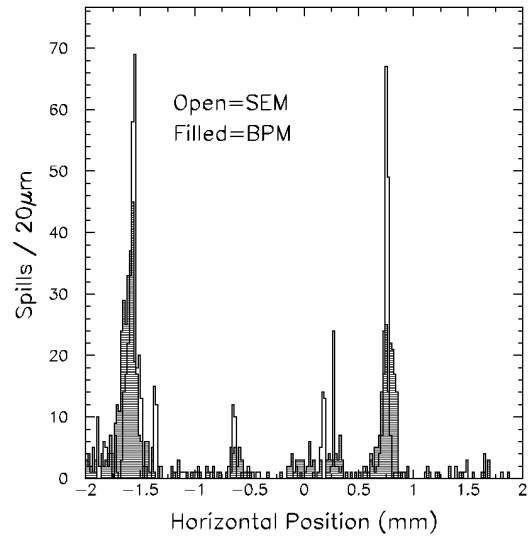

Figure 5: 1-dimensional projections of the data from the beam scan of Figure 3 onto the SEM and BPM axes.

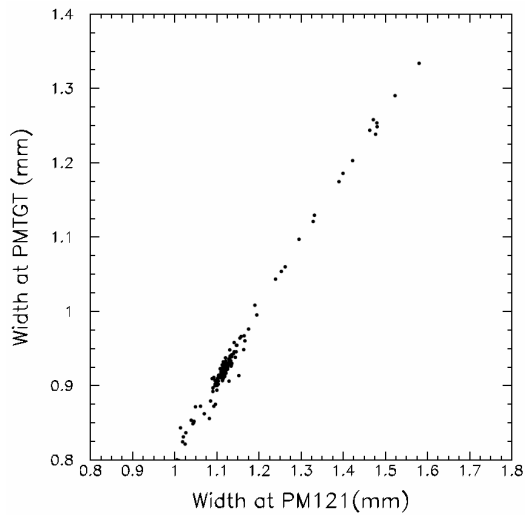

Figure 6: Beam width as observed at the last two profile monitor SEM's before the NuMI target.

Figure 8 shows the beam loss as measured by two ionization chamber loss monitors downstream of a pair of SEM chambers. The two chambers, PM117 and PM118, are both removed and then sequentially re-inserted into the beam. PM118 is a Tungsten-wire SEM (25µmØ), while PM117 is a Ti foil SEM (both are 1.0mm pitch). The relative increase in observed loss at both stations indicates that that PM118 causes approximately 5.9 times more loss than PM117. We would expect that the W-wire SEM would have a factor 6.7 more loss if scattering (radiation lengths) determined beam loss or a factor 1.9 more loss if nuclear interactions determined the loss in the SEM material. That our measurement is somewhat in between these two values indicates that beam loss is due in part to both these two effects. These measurements indicate that the foil SEM's reduce the beam loss relative to the 75µmØ W wire SEM's [4] by a factor ~50-60.

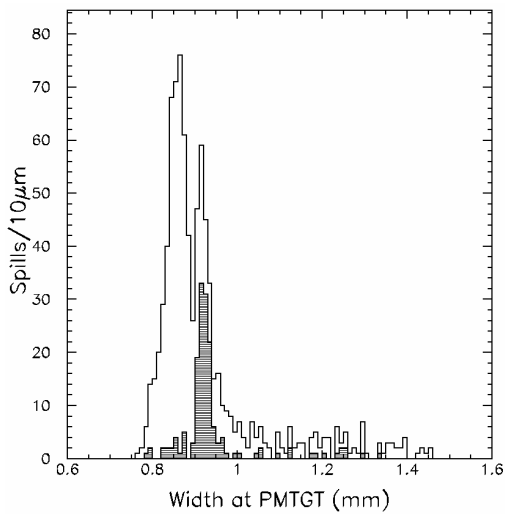

Figure 7: Projection of the data in Figure 5 onto the PMTGT axis. Each entry in the histograms is one beam spill. The open (closed) histogram is the beam width observed over a several hour (~ ½ hr.) period.

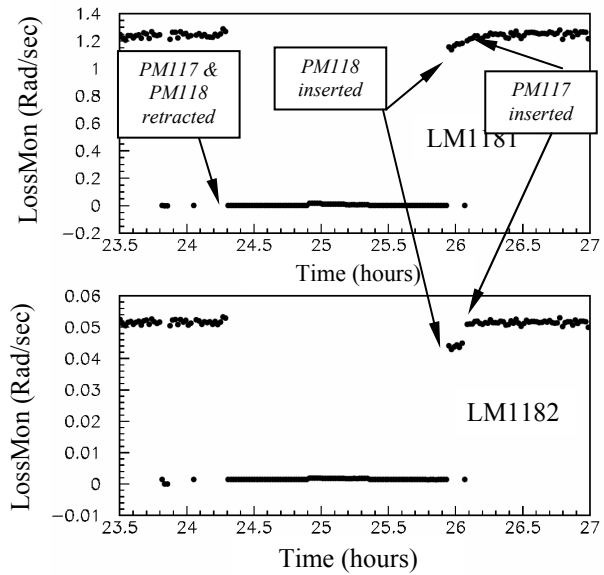

Figure 8: Beam loss at two stations LM1181 and LM1182 downstream of a pair of profile monitor SEM's PM117 and PM118. During the time period shown, PM117 and PM118 are retracted from and reinserted into the beam.

## SUMMARY

We have developed a large aperture segmented foil SEM for use in the 8 GeV and 120 GeV beam lines at FNAL. The foil SEM's are observed to significantly reduce beam loss, have satisfactorily performed up to beam intensities up to $2.5 \times 10^{13}$ ppp, and have beam width and centroid resolutions consistent with expectations.

## ACKNOWLEDGEMENTS

It is a pleasure to acknowledge G. Ferioli of CERN and J.D. Gilpatrick of LANL for many helpful conversations and sharing their experience of SEM grid design with us.